\documentclass[preprintnumbers,showpacs,amsmath,amssymb,prd,floatfix,twocolumn,superscriptaddress,nofootinbib]{revtex4}
\usepackage{graphicx}
\usepackage{epsfig}
\usepackage{bm}
\usepackage{amsfonts}

\usepackage{color}

\def\beq{\begin{equation}}
\def\eeq{\end{equation}}

\def\be{\begin{equation}}
\def\ee{\end{equation}}
\def\bea{\begin{eqnarray}}
\def\eea{\end{eqnarray}}

\newcommand{\gsim}{\lower.7ex\hbox{$\;\stackrel{\textstyle>}{\sim}\;$}}
\newcommand{\lsim}{\lower.7ex\hbox{$\;\stackrel{\textstyle<}{\sim}\;$}}

\begin{document}

\title{Entropic cosmology: a unified model of inflation and late-time acceleration}

\author{Yi-Fu Cai}
\email{caiyf@ihep.ac.cn}
\author{Jie Liu}
\email{liujie@ihep.ac.cn}
\author{Hong Li}
\email{hongli@ihep.ac.cn}

\affiliation{ Institute of High Energy Physics, Chinese Academy of
Sciences, P.O. Box 918-4, Beijing 100049, P.R. China }

\affiliation{ Theoretical Physics Center for Science Facilities
(TPCSF), Chinese Academy of Sciences, P.R. China }

\begin{abstract}
Holography is expected as one of the promising descriptions of
quantum general relativity. We present a model for a cosmological
system involving two holographic screens and find that their
equilibrium exactly yields a standard Friedmann-Robertson-Walker
universe. We discuss its cosmological implications by taking into
account higher order quantum corrections and quantum nature of
horizon evaporation. We will show that this model could give rise
to a holographic inflation at high energy scales and realize a
late-time acceleration in a unified approach. We test our model
from the SN Ia observations and find it can give a nice fit to the
data.
\end{abstract}


\maketitle

\newpage

\section{Introduction}

Einstein's classical general relativity is commonly acknowledged as
the theory of gravitational interactions for distance sufficiently
large compared to the Planck length. This validity has become the
foundation of modern cosmology in describing the dynamics of our
universe. However, its quantum effects are expected to become
important at high energy scale, namely at very early time of
cosmological evolution. Especially, the quantization of Einstein
gravity has long been known to be perturbativly nonrenormalizable.
Various attempts on solving this issue have been intensively studied
in the literature. It is widely believed that the quantization of
Einstein gravity is related to the solution to Big Bang singularity
of our universe.

As early as the discovery of black hole thermodynamics by
Bekenstein\cite{Bekenstein:1973ur} and
Hawking\cite{Hawking:1974sw}, people have realized that a
nonperturbative feature of Einstein gravity may be related to the
holographic thermodynamics. Especially, 't Hooft proposed the
holographic principle as a particular property of quantum gravity
which states that the description of a volume of space can be
thought of as encoded on a boundary of this system, preferably a
light-like boundary like a gravitational
horizon\cite{'tHooft:1993gx}. Subsequently, this issue is
extensively discussed in cosmology\cite{Gibbons:1977mu,
Bousso:2002ju} and recently realized in the context of
developments in string theory\cite{Maldacena:1997re}. Therefore,
it provides a promising description of quantum general relativity.
An extended holographic picture was conjectured by Verlinde and in
this scenario Einstein gravity is originated from an entropic
force arising from the thermodynamics on a holographic
screen\cite{Verlinde:2010hp}(see also \cite{Padmanabhan:2009vy,
Padmanabhan:2009kr, Padmanabhan:2010xh} and references therein for
earlier studies along this direction). In this scenario, however,
there exists a key controversial issue whether gravity is
fundamental or emergent\cite{Gao:2010yy, Culetu:2010ua}, and thus
relevant reinterpretations of Verlinde's Entropic force was
discussed in \cite{Hossenfelder:2010ih}.

Recently, a much explicit formulation of Entropic gravity theory
was suggested by Easson, Frampton and Smoot (EFS) , in which the
general relativity is still a fundamental theory but including a
boundary term. In this picture the holographic entropic force
arises from the contribution of boundary
terms\cite{Easson:2010av}. This model was soon applied to realize
current acceleration \cite{Easson:2010av} and inflationary period
at early universe \cite{Easson:2010xf}(see \cite{Wang:2010jm} for
a study of entropic inflation within Verlinde's proposal). In both
Verlinde's proposal and EFS one, one should be very careful of
whether they can explain the current cosmological observations
consistently\cite{Danielsson:2010uy}. First of all, let us assume
gravity is entropic, when applied into cosmology, we should
explain why our CMB radiation and the holographic screen are not
in thermal equilibrium. This problem is very manifest, since the
temperature of our CMB radiation and thus our universe is observed
as $T_{CMB}=2.73$K, but the horizon temperature can be easily
estimated as $T_H\sim{O}(10^{-30})$K. In such a thermal system out
of equilibrium, the heat transfer ought to be very strong.
Therefore, if such a heat transfer occurs today and leads to an
acceleration, then our universe is always accelerating in the past
without radiation and matter dominations since the temperature gap
is always very large at early times (we have
$T_H\sim{T_{CMB}^2}\sqrt{G}$ and therefore this system can only be
in thermal equilibrium at Planck scale). This conclusion explains
why the modified Friedmann equation appeared in the EFS paper is
so different from the normal one in Einstein gravity.

Concerning the above question, in the present work we are interested
in how to recover a standard Friedmann equation from an entropic
cosmological system. We suggest that there exist two holographic
screens with one being the approximate Hubble horizon which is
similar to the de-Sitter (dS) horizon, while the other Schwarzschild
horizon. Each screen has its own thermodynamics which is independent
of that of the other. Consider the classical dynamics of such a
cosmological system, we find that its thermal equilibrium
corresponds to a standard FRW universe. Therefore this model is able
to explain the normal thermal history observed in our universe.
Moreover, we consider quantum corrections to the area entropy and
obtain an EFS universe a little bit deviating from thermal
equilibrium. We study the cosmological implications of this model,
and find that it can drive an entropic expansion at early universe
and realize the late-time acceleration in a unified approach.

The letter is organized as follows. In section II we briefly review
the idea of entropic force from the viewpoint of an effective action
description. In section III we provide an explicit formulation of
the FRW universe from the classical thermal equilibrium state of
double holographic screens. Effects from higher order quantum
corrections of the holographic entropy and a quantum evaporation of
the inner horizon are studied in this model. In section IV we study
the cosmological implications of this model involving quantum
corrections. Our results show that in this model a holographic
inflation could be obtained at high energy scales, and a quantum
evaporation process of the inner horizon is related to the
realization of the late-time acceleration. We confront this model
with the latest observations at the end of this section, and the
results are in agreement with observational constraints. Section V
presents a summary and discussions of the related works. We take the
convention $c=k_B=\hbar=1$ in this letter.

\section{Review of entropic force and effective action description}

We start with a discussion on the standard approach to studying
quantum field theory. The viewpoint of modern physics suggests any
fundamental theories can be described by an effective action at
certain energy scale. So does Einstein gravity. Under this
assumption, we can obtain the equations of motion to describe the
dynamics of these theories from a variational principle. In a
usual quantum field theory, we can integrate out the boundary
terms which do not change its physics in Minkowski spacetime. When
dealing with a gravitational system, however, one should be very
careful of the boundary terms which may play an important role of
the background evolution.

In general, the effective action of a gravitational system
including matter fields and surface terms is described by
\begin{eqnarray}
 {\cal I} = \int_M\bigg(\frac{R}{16\pi G}+{\cal L}_m\bigg) +\oint_{\partial M}{\cal L}_b~,
\end{eqnarray}
where $R$ is the Ricci scalar of the whole spacetime, ${\cal L}_m$
is the Lagrangian of matter fields living in the bulk, and ${\cal
L}_b$ is the corresponding Lagrangian describing the physics of
the boundary. Clues from string theory and AdS/CFT indicate that
the boundary terms should include the extrinsic curvature of the
boundary and holographic dual gauge theories. In this Letter, we
take this action as our starting point to describe a holographic
picture of the evolution of our universe.

By varying the action with respect to the metric, we can obtain the
Einstein field equation as follows,
\begin{eqnarray}
 R^{\mu\nu}-\frac{1}{2}Rg^{\mu\nu}
  = 8\pi GT_m^{\mu\nu} + J_b^{\mu\nu}~,
\end{eqnarray}
in which the last term $J_b$ is a current describing the exchange
of energy and momentum between the bulk and the boundary. In
usual, it is described by a delta function in order to satisfy the
locality of the theory. However, in the frame of a holographic
picture, this term is determined by the holographic description of
boundary physics and so is a nonlocal effect, which in our letter
corresponds to an entropic force in the universe\footnote{We thank
Yi Wang for pointing out this issue on the Buzz discussion.}.

Now we assume the boundary physics can be described by
thermodynamics satisfying a holographic distribution. Therefore,
the number of degrees of freedom on this holographic screen is
proportional to its area which takes $N\propto A$. In this case,
the classical holographic entropy on this screen is given by
\begin{eqnarray}\label{Sb}
 S_b=\frac{A}{4G}=\frac{\pi}{G} r_b^2~,
\end{eqnarray}
where $r_b$ is the radius location of the boundary surface. The
entropic force is determined by the variation of energy with
respect to the radius,
\begin{eqnarray}\label{Fe}
 F_{e}=-(\frac{dE}{dr})_b=-(T\frac{dS}{dr})_b=-\frac{2\pi}{G} T_br_b~,
\end{eqnarray}
in which $T_b$ is the temperature of the boundary system and we
have applied Eq. (\ref{Sb}) in the above formula. According to the
Unruh effect, when a test particle with mass $m$ is located near
by the horizon, the variation of the entropy on this horizon with
respect to the radius takes the form of
\begin{eqnarray}\label{Unruh}
 \frac{dS}{dr}=-2\pi m~,
\end{eqnarray}
and thus the combination of Eqs. (\ref{Fe}) and (\ref{Unruh})
yields an entropic acceleration $a_e$ as follows,
\begin{eqnarray}\label{ae}
 a_e\equiv\frac{F_e}{m}=2\pi T_b~.
\end{eqnarray}
The corresponding entropic pressure takes a negative form
$P_e=F_e/{A_b}=-{T_b}/{2Gr_b}$, and so is expected to drive the
current acceleration of our universe.

Now we apply the above results into a homogeneous and isotropic flat
Friedmann-Robertson-Walker (FRW) universe described by
\begin{eqnarray}
ds^2=dt^2-a(t)^2dx^idx^i~,
\end{eqnarray}
and the Einstein field equation gives the acceleration equation
for the scale factor as follows,
\begin{eqnarray}
 \frac{\ddot a}{a}=-\frac{4\pi G}{3}(\rho+3p)+\frac{a_e}{L_b}~,
\end{eqnarray}
where $L_b$ is a length scale relevant to the location of the
holographic screen, of which a natural choice is to near by the
Hubble horizon\footnote{This horizon is not necessary the exact
form of Hubble horizon. So we introduce an undetermined
coefficient $\beta$ which will constrained by cosmological
observations.}
\begin{eqnarray}
 L_b=r_H=\frac{1}{\beta H}~,
\end{eqnarray}
with the horizon temperature being
\begin{eqnarray}
 T_H=\frac{\beta H}{2\pi}~,
\end{eqnarray}
and $H\equiv \dot a/a$ is the Hubble parameter of the universe.
Finally, we arrive at a modified Friedmann acceleration equation
as follows,
\begin{eqnarray}
\frac{\ddot a}{a}=-\frac{4\pi G}{3}(\rho+3p)+\beta^2H^2~.
\end{eqnarray}
From this equation we can learn that, if the coefficient $\beta^2$
is of order $O(1)$, the Friedmann equations describing the
evolution of our universe would be changed too much by taking into
account such a holographic screen, which is hardly able to recover
the usual form in radiation and matter dominated periods. This
point also appeared in Ref. \cite{Easson:2010av}, and the authors
of that paper suggest to modify the coefficients before the
temperature and introduce an alternative entropic acceleration.

\section{Formulating FRW universe from thermal equilibrium of double holographic screens}

The key reason leading to the above over-modified Friedmann equation
is that in this gravitational system we have only considered a
single holographic screen related to the Hubble horizon. In this
case, there exists a temperature gap between the bulk universe (CMB
temperature) and the horizon, which indicates that the whole
gravitational system is not in thermal equilibrium and all the
particles in the bulk universe will fall down onto the horizon due
to such an unbalanced effect.

\subsection{A delicate picture of double holographic screens}

In such a cosmological system, one another holographic screen ought
to be taken into account, i.e., the Schwarzschild horizon. This
horizon is motivated by physics of black holes, which is another
promising approach to understanding the quantum information of
Einstein's gravity. A black hole has many remarkable properties,
namely, the association with thermodynamics and the reflection of
holography introduced at the beginning of the current Letter. The
simplest black hole solution in four dimensions corresponds to a
Schwarzschild spacetime discovered many years ago. In this solution,
the whole geometry is divided into two causal independent regions by
an event horizon at the radius
\begin{eqnarray}
 r_S=2GM=2G\int_M\rho dV=\frac{8\pi G\rho}{3\beta^3H^3}~.
\end{eqnarray}
Its corresponding temperature is given by
\begin{eqnarray}
 T_S=\frac{1}{8\pi G M}=\frac{3\beta^3H^3}{32\pi^2G\rho}~,
\end{eqnarray}
and therefore, according to the relationship obtained in Eq.
(\ref{ae}), we obtain a modified entropic acceleration
\begin{eqnarray}
 a_e=2\pi (T_H-T_S)
  =\beta H\bigg(1-\frac{3\beta^2H^2}{16\pi G\rho}\bigg)~,
\end{eqnarray}
which reflects a competition of entropic effects from the outer
horizon and the inner horizon. Interestingly, if these two
holographic screens are delicately on thermal equilibrium with
$T_H=T_S$ and we choose the coefficient $\beta=\sqrt{2}$, one can
recover the exact form of the traditional Friedmann equation
\begin{eqnarray}\label{FRW1}
 H^2=\frac{8\pi G}{3}\rho \bigg|_{\rm equilibrium}~,
\end{eqnarray}
which coincides with the normal Friedmann equation. In this case
one can check the Schwarzschild radius is smaller than the Hubble
radius. Moreover, the continuous equation for matter fields in the
bulk spacetime satisfies
\begin{eqnarray}\label{FRW2}
 \dot\rho+3H(\rho+p)=0~,
\end{eqnarray}
when there is no coupling to the boundary surfaces. These two
equations of motion are self-complete and yields the normal
evolution of our universe.

\begin{figure}[htbp]
\includegraphics[scale=0.5]{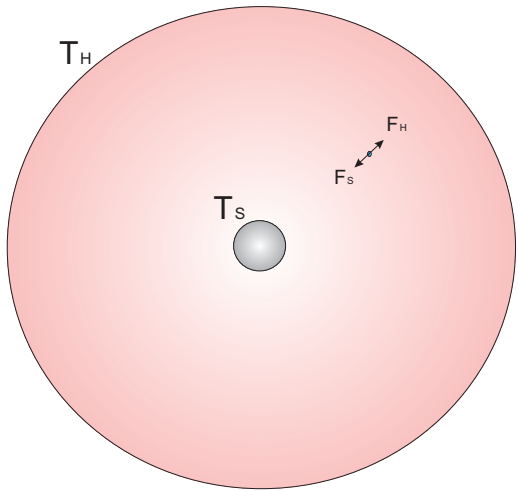}
\caption{A sketch of the dynamics of a test particle under double
holographic screens. The inner screen is the Schwarzschild horizon
and the outer one is comparable to the Hubble horizon. The
particle will feel two opposite entropic forces and finally fall
into a delicate balanced state when two screens are on thermal
equilibrium. } \label{Fig:sketch}
\end{figure}

In the following we depict this holographic picture with a cartoon
sketch as shown in Fig. \ref{Fig:sketch}. In this figure, one can
see that any matter fields in the bulk spacetime would feel two
different entropic forces of which the directions are opposite.
Therefore, the energy stored in the bulk spacetime will transfer to
the holographic screen with lower temperature until two temperatures
equal to each other. Once $T_H=T_S$, the universe will arrive at a
delicate balanced state and evolves according to its own equations
of motion without energy exchanges among these screens.
Additionally, during the whole evolution of the universe the second
law of thermodynamics is preserved.

\subsection{A modified Friedmann equation with quantum corrections}

Up to now, we have studied the holographic description of an FRW
universe with double screens in classical level. However, it has
been studied for many years in the field of quantum Einstein gravity
and string theory that, the form of a holographic entropy should be
improved when higher order quantum corrections are taken into
account\cite{Easson:2010xf}. Namely, both the number of string
states\cite{Strominger:1996sh, Halyo:1996xe, Solodukhin:1997yy} and
the holographic renormalization group flow\cite{Alvarez:1998wr}
yields an improved form of the entropy with leading order
corrections as follows,
\begin{eqnarray}
 S_i=\frac{1}{4G}(A_i+g_iG\ln\frac{A_i}{G}+...)~,
\end{eqnarray}
where $i=S,H$ denotes the inner and outer horizon respectively. In
the above formula, the coefficient $g_i$ is determined by specific
environment and here we would like to leave it as a free
parameter.

In this case, the entropy force of a holographic screen is changed
to be
\begin{eqnarray}
 F_e=-T\frac{dS}{dr}=-\frac{2\pi}{G} r T(1+\frac{gG}{4\pi r^2}+...)~,
\end{eqnarray}
at leading order. With double holographic screens, we obtain an
improved acceleration equation, which is expressed as
\begin{eqnarray}\label{HFRW1}
 \frac{\ddot a}{a} = -\frac{4\pi G}{3}(\rho+3p) + f(\rho, H)~,
\end{eqnarray}
with the form of surface function being
\begin{eqnarray}
 f(\rho, H)
  &=& \beta^2H^2\bigg(1-\frac{3\beta^2H^2}{16\pi G\rho}\bigg) \nonumber\\
  &+& \frac{g_HG\beta^4H^4}{4\pi}(1-\frac{27g_S\beta^6H^6}{1024g_H\pi^3G^3\rho^3})+...~.
\end{eqnarray}

\subsection{Dissipating equation of the Schwarzschild horizon}

In the physics of quantum gravity, there is an important lesson
from holography, i.e., when there exists Hawking radiation, the
boundary horizon loses a small amount of its
energy\cite{Hawking:1974sw}. This can be achieved similar to the
well-known studies of a massive primordial black
hole\cite{Carr:1974nx} (see the recent study
\cite{Custodio:1998pv, Jacobson:1999vr, Bean:2002kx,
Harada:2004pf}). In a generic cosmological background, its
dissipating equation is given by\cite{Bean:2002kx},
\begin{eqnarray}\label{dissipate}
 \frac{dM}{dt} = -\frac{\alpha}{M^2}+\kappa\frac{M^2}{t^2}~.
\end{eqnarray}
Here the $\alpha$ term describes the evolution of the
Schwarzschild horizon via Hawking radiation, and so
$\alpha=1/15360\pi G^2$. Moveover, the $\kappa$ term denotes the
accretion process and the value of $\kappa$ is determined by the
background cosmological evolution at initial moment, which in a
usual case corresponds to a critical point $t_{C}$ when the whole
system is on the thermal equilibrium and the standard Friedmann
equation is recovered. Thus one gets $\kappa=16\rho_{C}t_{C}^2$.

The dissipating equation (\ref{dissipate}) can not be solved in an
analytic approach. However, one may notice that, for such a
horizon evolving at late time of the universe, the $\kappa$ term
is strongly suppressed by the age of our universe. As a
consequence, one can solve the dissipating equation and obtain an
evaporation time of the form
\begin{eqnarray}\label{tauS}
 \tau_S \simeq \bigg(\frac{M_C}{10^{15}g}\bigg)^3 \times 10^{17}s~,
\end{eqnarray}
where $M_C$ denotes the Schwarzschild mass at the critical moment.
From this relation, we find that in order to let the age of such a
Schwarzschild horizon in the order of our universe, there has to
be $M_C\sim O(10^{38}{\rm GeV})$ at the critical moment. We leave
the explicit values of the coefficients to be determined or
constrained by astronomical observations which will be discussed
later. Note that if this horizon has not evaporated out completely
until today, in principle it could be expected to be observable by
CMB experiments. Namely, the radiation arising from the
evaporation of the inner horizon could bring an amplification of
CMB spectrum at sub-Hubble scales. This issue deserves a study in
detail in the future\cite{limiao}.

In addition, we would like to comment on the quantum decay of the
outer horizon. Since this horizon is similar to the event horizon
of the dS spacetime, according to the work of Coleman and De
Luccia\cite{Coleman:1980aw}, the decay time of a meta-stable dS
vacuum has an approximate expression $\tau_H\sim {\cal P}_H^{-1}$.
By neglecting all the sub-exponential factors, we have
\begin{eqnarray}
 {\cal P}_H \sim e^{-{\pi}/{GH^2}}~,
\end{eqnarray}
which is exponentially suppressed by the age of the universe.
Therefore, we can neglect the radiation emission of the outside
horizon throughout the whole cosmological evolution.

\section{Cosmological implications}

We now study the cosmological implications of double holographic
screens. Specifically, we discuss realizations of the late-time
acceleration and entropic evolution of early time universe
respectively. We find there might be a potential way to unify these
two processes together in the frame of entropic cosmology.

\subsection{Early time evolution of the holographic universe}

According to the well-known knowledge of thermodynamics, a system
usually evolves from an unstable state to the equilibrium. This
process is described by a generalized acceleration equation appeared
in Eq. (\ref{HFRW1}), but we still need one another equation of
motion to solve this system self-complete. We consider the
conservation of the whole energy in this system, which sums up the
contributions from the double holographic horizons and the universe,
which can be described by\footnote{We notice our conservation
equation is manifestly different from Eq. (4) in
\cite{Danielsson:2010uy} and consequently the second Friedmann
equation is no longer in form of $\dot H=-4\pi G(\rho+p)$. }
\begin{eqnarray}
 \Delta E_{u}+\Delta E_{S}+\Delta E_{H}=-p\Delta V~,
\end{eqnarray}
where $\Delta E_i=T\Delta{S}_i$ for $i=S,H$ respectively. This
yields an improved continuous equation
\begin{eqnarray}\label{HFRW2}
 \dot\rho+3H(\rho+p)=\Gamma~,
\end{eqnarray}
with an effective coupling term $\Gamma$ being
\begin{eqnarray}
 \Gamma = \frac{27\beta^6H^6}{1024\pi^3G^3\rho^3}\dot\rho
  +\frac{3\beta^2H\dot{H}}{4\pi G}
  \bigg(1-\frac{27\beta^4H^4}{256\pi^2G^2\rho^2}\bigg)~,
\end{eqnarray}
in classical level. One can check when $\beta=\sqrt{2}$ and
$T_H=T_S$, the coupling $\Gamma$ vanishes and Eq. (\ref{HFRW2}) is
in agreement with the normal continuous equation (\ref{FRW2}).

Consider a radiation dominated universe at high energy scales. We
assume the equation of state for the radiation in the universe is
determined by its intrinsic physical nature which gives
$p=\rho/3$. To combine the equations of motion (\ref{HFRW1}) and
(\ref{HFRW2}) and only keep the leading order quantum correction
to the entropy, one can approximately solve out the following
solution for the Hubble parameter
\begin{eqnarray}\label{HFRW}
 H^2 = \frac{8\pi G}{3}\bigg[\rho+\frac{8(g_H-4g_S)}{69}G^2\rho^2+...\bigg]~,
\end{eqnarray}
at early universe. During the semi-analytical derivation, we find
again that $\beta$ is required to be $\sqrt{2}$ approximately in
order to make sure the above calculation self-consistent, which
indicates the system would approach to the thermal equilibrium
state along with the cosmological evolution.

From Eq. (\ref{HFRW}), one can see that a standard Friedmann
equation can be achieved if $g_H=4g_S$ even without a thermal
equilibrium at early times. There exist two branches of
cosmological solutions at high energy scales. The first one
describes an expanding universe with $g_H>4g_S$, and the Hubble
parameter is proportional to the energy density at high energy
scales. In this branch the $\rho^2$ term could make the early time
inflation much easier be realized, so would provide an implement
of holographic inflation as studied in \cite{Easson:2010xf,
Wang:2010jm}. In the following we will illustrate its possibility
in the example of a radiation dominated universe.

In the case of $g_H\neq4g_S$, the Hubble parameter will be
dominated by the second term in the rhs of Eq. (\ref{HFRW}) when
the energy density of the universe reaches a critical value
\begin{eqnarray}
 \rho_C\simeq\frac{69}{8(g_H-4g_S)G^{2}}~,
\end{eqnarray}
for a radiation dominated entropic universe. One can solve the
equations of motion and find that the energy density of the
universe evolves as
\begin{eqnarray}\label{inflation1}
 \rho\simeq\sqrt{\rho_C^2-\frac{512\pi^3t}{27G^{9/2}(g_H-4g_S)^{5/2}}}~,
\end{eqnarray}
where we set the initial moment of the entropic inflation
$t_i\rightarrow -\infty$ and choose $t_C=0$ corresponds to the
beginning moment of normal Friedmann equation. To proceed, we
obtain the form of Hubble parameter as follows,
\begin{eqnarray}
 H(t) \simeq 24.25\times\frac{(-t)^{1/2}}{[G(g_H-4g_S)]^{3/4}}~,
\end{eqnarray}
when $|t|\gg1$. In order to characterize the inflationary process,
it is convenient to introduce the slow-roll parameter $\epsilon$,
which is in form of
\begin{eqnarray}
 \epsilon \equiv -\frac{\dot H}{H^2}
 \simeq 2.06\times10^{-2}\frac{[G(g_H-4g_S)]^{3/4}}{(-t)^{3/2}}~.
\end{eqnarray}
This parameter can be much less than unity when
$|t|\gg\sqrt{G(g_H-4g_S)}$. Consequently, we obtain a period of
holographic inflation.

Note that, if we assume the critical scale obtained above
corresponds to the critical mass appeared in Eq. (\ref{tauS}), it
requires $\rho_C$ to be around or less than $O(10^{-3}{Mp})^4$
with $M_p\equiv1/\sqrt{G}$. Therefore, in order to let the inner
horizon evaporate within the age of our universe, one expects that
the value of $|g_H-4g_S|$ should be finely tuned as large as
$O(10^{12})$. As an explicit example, we choose
$|g_H-4g_S|=10^{16}$ and thus get the efolding number
\begin{eqnarray}
 {\cal N}\equiv\int^tH(t)dt\simeq1.62\times10^{-11}(-M_pt)^{3/2}~.
\end{eqnarray}
where we introduce the Planck time $t_p=\sqrt{G}$.

\begin{figure}[htbp]
\includegraphics[scale=0.8]{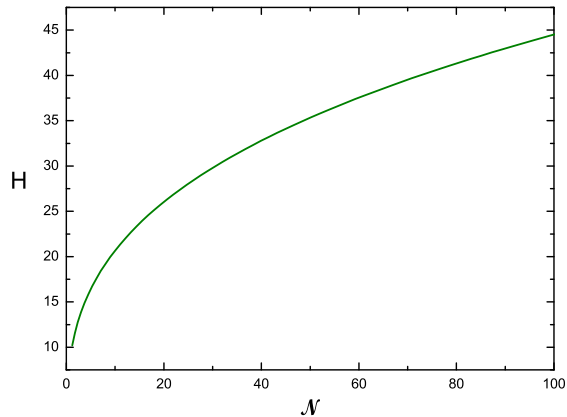}
\caption{A plot of the evolution of the Hubble parameter $H$ in
the scenario of holographic inflation as a function of the
efolding number ${\cal N}$. In the numerical computation, we
choose $g_S=0$ and $g_H=10^{16}$. The Hubble parameter is in unit
of $10^{-8}M_p$. } \label{Fig:Hubble}
\end{figure}

\begin{figure}[htbp]
\includegraphics[scale=0.8]{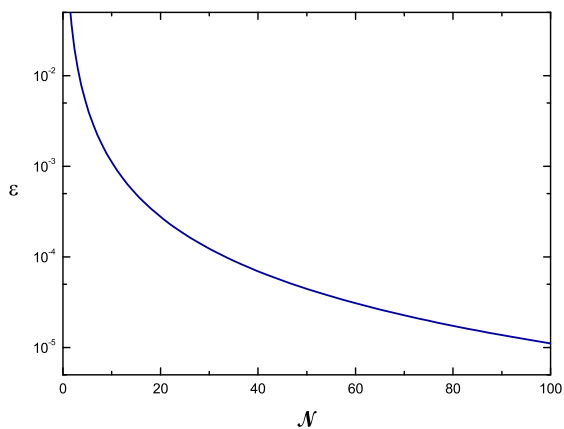}
\caption{A plot of the evolution of the slow-roll parameter
$\epsilon$ in the scenario of holographic inflation as a function
of the efolding number ${\cal N}$. In the numerical computation,
we choose $g_S=0$ and $g_H=10^{16}$. } \label{Fig:slow}
\end{figure}

We plot the evolution of the Hubble parameter $H$ and the
slow-roll parameter $\epsilon$ as functions of the efolding number
${\cal N}$ in Figs. \ref{Fig:Hubble} and \ref{Fig:slow}. From
these figures, one can read that for ${\cal N}=60$ there is
$\epsilon\simeq3\times10^{-5}$ and $H=3.8\times10^{-7}M_p$.
Finally, we find that in the model of double holographic screens
the universe could experience an exponential expansion at early
times due to the effect of entropic force.

The second branch corresponds to the case of $g_H<4g_S$. In this
case we expect there exists a cosmological bouncing solution which
may avoid the initial singularity. Since this topic is beyond the
aim of the current Letter, we would like to leave it in future
studies.

\subsection{Late-time acceleration and comparing with the SN Ia data}

Along with the decreasing energy scale, the universe will
gracefully exit from a primordial epoch either a normal expansion
or a holographic inflation, and then smoothly link with the
standard thermal history as we observed. During this period, our
universe has already arrived at the delicate thermal balanced
state with its evolution satisfying the standard Friedmann
equation. However, as mentioned in the above section, one should
take into account the horizon evaporation via Hawking radiation.
Therefore, the double holographic screens can also lead to the
late-time acceleration of the universe once the inner horizon
evaporates.

\begin{figure}[htbp]
\includegraphics[scale=0.5]{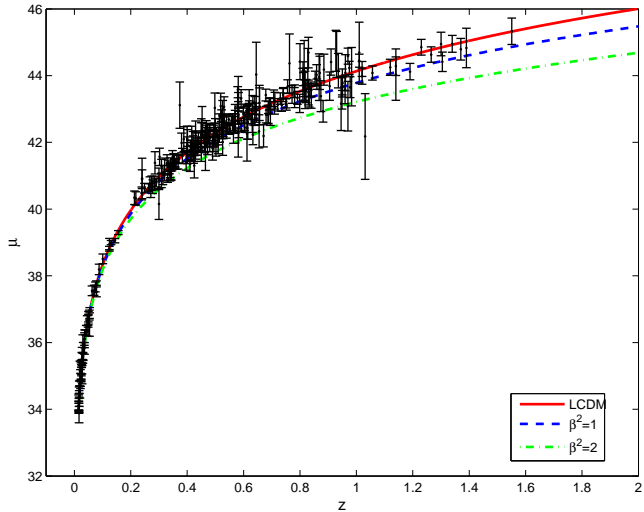}
\caption{A numerical plot of the Hubble diagram caused of entropic
acceleration and its comparison to $\Lambda$CDM. The red solid
line is given by $\Lambda$CDM model, the blue dashed line is given
by $\beta^2=1$ and the green line is from $\beta^2=2$. The black
dots with the error bars are the SN Ia "union" compilation
sample.} \label{Fig:dismod}
\end{figure}

The late-time acceleration can be realized via the equation of
motion (\ref{HFRW1}) where the form of $f$ approximately takes
$\beta^2H^2$ which coincides with the result obtained in Ref.
\cite{Easson:2010av}. However, we leave the coefficient $\beta$ to
be a free parameter due to lacks of detailed information of the
inner horizon evaporation. In principle, we can constrain the
model by fitting to the Type Ia Supernovae (SN Ia) data and give
the constraints on the model parameter $\beta$. In order to see
the rationality of the model transparently, we plot the distance
moduli given by our model with specific values of $\beta$ in Fig.
\ref{Fig:dismod} as well as the SN Ia data given by the "union"
compilation\cite{Kowalski:2008ez}. We use the metric theory of
gravity and the general formula of the luminosity distance is
given by:\be
d_L(z)=(1+z)\int_0^z\frac{dz^{\prime}}{H(z^{\prime})},\ee where
$z$ is the redshift define by $a_0/a=1+z$. In the $\Lambda$CDM
model, the Luminosity distance takes the
form\cite{Carroll:1991mt}:
\begin{eqnarray}
 d_L & = & c(1+z)H_0^{-1}|\Omega_k|^{-1/2}{\rm sinn}\{|\Omega_k|^{1/2} \nonumber \\
  & \times &
  \int_0^zdz[(1+z)^2(1+\Omega_Mz)-z(2+z)\Omega_\Lambda]^{-1/2}\}~, \nonumber \\
\end{eqnarray}
where $\Omega_k=1-\Omega_M-\Omega_\Lambda$, and ``sinn'' is
$\sinh$ for $\Omega_k > 0$ and $\sin$ for $\Omega_k < 0$. The blue
dashed line and the green line are given by $\beta^2=1,~2$
respectively, and the red solid line is given by the $\Lambda$CDM
model. From the plot, we can find a nice fit of the entropic
acceleration to the SN Ia data for the distance moduli given by
our model are well consistent with the data points at low
redshift. The idealistic case with $\beta^2=2$ deviates from the
observation at high redshift regime. This signature is
understandable since at high redshift regime the universe still
satisfies the standard Friedmann equation and the double
holographic screens are on thermal equilibrium as well. A more
explicit investigation on this issue should be addressed in future
in combination of the detailed study on dynamics of horizon
evaporations.

\section{Conclusion and Discussion}

Since awaken by Verlinde, the idea of entropic force has become an
important issue and its phenomenological applications were soon
considered in an FRW universe\cite{Cai:2010hk, Shu:2010nv,
Gao:2010fw, Zhang:2010hi, Wei:2010wwa, Ling:2010zc, Lee:2010bg},
and is expected to explain the current acceleration of our
universe\cite{Li:2010cj}, and drive an inflationary period at
early times\cite{Wang:2010jm}, and its effects on spherical
symmetric spacetime were discussed in \cite{Smolin:2010kk,
Caravelli:2010be, Wang:2010px, Myung:2010jv, Liu:2010na,
Cai:2010sz, Tian:2010uy, Jamil:2009qs, Jamil:2010xq}, and see
\cite{Myung:2010rz, Konoplya:2010ak, Paeng:2010dj, Kar:2010uy,
Hogan:2010zs, He:2010yf, Lee:2010ew, ChangYoung:2010rz,
Banerjee:2010yd, He:2010ct} for relevant discussions. The frontier
of modern physics suddenly goes back to pre-Einstein time one
hundred years ago.

In this Letter, we extended the picture of entropic cosmology and
suggested a scenario of double holographic screens to explain the
past thermal expansion of our universe. We also studied the
quantum signatures of this model motivated from physics of quantum
gravity, and specifically we considered the higher order quantum
corrections to the holographic entropy and the process of horizon
evaporations. We found that the higher order quantum corrections
to the entropic force may give rise to an implement of holographic
inflation. In the meanwhile, the evaporation of the inner horizon
could bring a realization of late-time acceleration. In order to
let this acceleration happen at current time, we found there
exists a fine-tuning to the coefficients of higher order
corrections. We test our model from the SN Ia observations, and
find it can give a nice fit to the SN Ia data. The unification of
inflation and dark energy era was earlier discussed in Refs.
\cite{Nojiri:2005sr, Nojiri:2005pu} by introducing a phantom
degree of freedom.

We would like to point out that the model we considered is still a
toy model with many detailed clues ignored. Among them the most
important issue is the study of primordial perturbations seeded by
statistic fluctuations on the holographic screens, since we expect
these thermal fluctuations could give rise to a nearly
scale-invariant spectrum so that explain the CMB observations. We
will perform a much complete and careful study on this issue in
near future.

At the end of this Letter, we would like to make a few comments on
the possibility of the avoidance of big bang singularity in
entropic cosmology. In the main text, we have studied the
cosmological implications of the model of double holographic
screens at early universe by considering the higher order quantum
corrections. Moreover, in quantum physics, there could be more
arguments supporting the avoidance of the Big Bang singularity.
Namely, as the Heisenberg uncertainty principle states, in quantum
theory a test particle is described by a wave packet, which moves
in the bulk spacetime. Consider the measurement of the absolute
position of this particle. It could be anywhere since the
particle's wave packet has non-zero amplitude, meaning the
position is uncertain. The Heisenberg uncertainty principle
requires,
 $\delta r \delta E \geq \frac{1}{2}~.$
To combine the above uncertainty relation and Eq. (\ref{Fe}), one
can obtain a minimal length scale for the thermal system
\begin{eqnarray}
\Delta r\simeq\sqrt{\frac{G}{4\pi Tr}}\sim l_{pl}~,
\end{eqnarray}
which implies that our universe cannot be shrunk into
trans-Planckian scale. Therefore, we may get a nonsingular cosmic
evolution of the universe at early times. In this case, we expect
the model could realize the late-time acceleration and also avoid
the initial big bang singularity in a unified approach without
quantum instability\footnote{It is widely noticed that in the
frame of standard Einstein gravity, a nonsingular bounce model
offer suffers from quantum instability due to a ghost degree of
freedom\cite{Cai:2007qw, Cai:2007zv}, this statement can also be
extended into the cyclic cosmology\cite{Cai:2009zp, Cai:2006dm,
Xiong:2007cn, Xiong:2008ic}. Recently, a nonsingular bounce model
was achieved in the frame of nonrelativistic gravity
theory\cite{Brandenberger:2009yt, Cai:2009in, Cai:2009hc}(see also
\cite{Calcagni:2009ar, Kiritsis:2009sh}).}. If this model has a
matter dominated contraction, it was found that both the thermal
\cite{Cai:2009rd} and quantum \cite{Wands:1998yp, Finelli:2001sr,
Cai:2008qw, Starobinsky:1979ty} fluctuations are able to provide a
scale-invariant spectrum with local featured
signatures\cite{Cai:2008qb, Cai:2008ed} and sizable
non-Gaussianities\cite{Cai:2009rd, Cai:2009fn}, which may be
responsible for the current cosmological observations. We note the
study of entropic force with Heisenberg uncertainty principle in
the original Verlinde's model appeared in \cite{Vancea:2010vf},
and its extended form was analyzed in \cite{Zhao:2010vt,
Lee:2010fg, Kuang:2010gs, Ghosh:2010hz, Munkhammar:2010rg,
Modesto:2010rm}.

\section*{Acknowledgements}

We wish to thank Robert Brandenberger, Damien Easson, Xian Gao,
Miao Li, Stone Pi, Yun-Song Piao, Taotao Qiu, Emmanuel N.
Saridakis and Yi Wang for stimulating discussions. We especially
thank Prof. Xinmin Zhang for extensive support and valuable
suggestions. Our work is supported in part by the National Science
Foundation of China under Grants No. 10533010 and 10803001 and the
Youth Foundation of the institute of high energy physics under
Grant Nos. H95461N.

\end{document}